# Patterns of Selection of Human Movements I: Movement Utility, Metabolic Energy, and Normal Walking Gaits

**Stuart Hagler**
Oregon Health & Science University
Portland, OR, USA
haglers@ohsu.edu

**Abstract:** The biomechanics of the human body allow humans a range of possible ways of executing movements to attain specific goals. Nevertheless, humans exhibit significant patterns in how they execute movements. We propose that the observed patterns of human movement arise because subjects select those ways to execute movements that are, in a rigorous sense, optimal. In this project, we show how this proposition can guide the development of computational models of movement selection and thereby account for human movement patterns. We proceed by first developing a movement utility formalism that operationalizes the concept of a best or optimal way of executing a movement using a utility function so that the problem of movement selection becomes the problem of finding the movement that maximizes the utility function. Since the movement utility formalism includes a contribution of the metabolic energy of the movement (maximum utility movements try to minimize metabolic energy), we also develop a metabolic energy formalism that we can use to construct estimators of the metabolic energies of particular movements. We then show how we can construct an estimator for the metabolic energies of normal walking gaits and we use that estimator to construct a movement utility model of the selection of normal walking gaits and show that the relationship between avg. walking speed and avg. step length predicted by this model agrees with observation. We conclude by proposing a physical mechanism that a subject might use to estimate the metabolic energy of a movement in practice.

1 Introduction

Due to the complexity of human biomechanics (e.g. articulated limbs), the execution of most human movements requires sophisticated, coordinated sensory-motor and cognitive processes. [1, 2] Although most human movements can be executed using a range of trajectories or orbits consistent with the biomechanical constraints, human movements exhibit significant patterns across and within individual subjects (e.g. there is a relatively narrow range of bipedal locomotor movements that would be considered normal human walking gait). Deviations from normal movement patterns have therefore been used by clinicians to assess cognitive, neural, muscular, and skeletal health. Characterizations of normal and abnormal movements can also be used to infer aspects of cognitive and neuromuscular processes. The goal of this project is to investigate the principles that constrain human movements into recognizable patterns. The general idea is that a movement trajectory is the result of an optimization process under a combination of constraints such as efficiency, safety, etc.

One optimization principle that researchers have used in the past to analyze movement has been energy minimization. [3-9] Two types of energy can be associated with a movement: (i) the *mechanical energy* of the trajectory of the body in space, and (ii) the *metabolic energy* of the sequence of muscle activations that move the body (i.e. the metabolic energy above the basal metabolic rate during the movement). The mechanical energy of the movement can be measured by observing the trajectories of the segments of the body, while the metabolic energy must be inferred by measuring the volumes of $O_2$ and/or $CO_2$ exhaled during respiration. The relationship between these two energies is not straight-forward; the metabolic energy of a movement is the total of the metabolic energy expended to generate all the muscle forces and give the body the required mechanical energy. [10] Due to the close relationship between mechanical and metabolic energy, some researchers have tried to characterize movement patterns using minimization of the mechanical energy, [3, 11, 12] while others have looked at modeling muscle to arrive at the metabolic work done when muscle is activated. [5, 6, 11, 13, 14] However, the latter approach involves complicated models that describe the behaviors of individual muscles. An alternative approach taken by Minetti et al. [15] is to derive an empirical metabolic energy function for the movement.

However, characterization of movement patterns solely using energy minimization seems insufficient to describe the full complexity of the phenomena. We can illustrate this by looking at walking gait in particular. While subjects do tend to walk in a way that minimizes the net metabolic energy given the walking speed, [3, 11, 12] other factors affect how a subject walks. In general, walking gait has also been shown to provide valuable medical information about subject cognitive and physical performance. [16] More particularly, studies have shown a relationship between walking speed and cognition. [17, 18] Walking speed has been found to decline as adults age, [19, 20] has been shown to be associated with survival in older adults. [21, 22] Walking speed tests can be used as predictors of adverse results related with health in older adults, [23] walking speed has been shown to be a quantitative estimate of risk of future hospitalization, [24] slower walking speed has been demonstrated in dementia patients compared to controls [25] and has been shown to precede cognitive impairment [26] and dementia, [27] the slowing of walking speed appears to take place secondary to the slowing of processing speed in the path leading to dementia. [28] Moreover, while subjects tend to move so as to minimize the net metabolic energy, [3, 11, 12] they may often choose to make movements with somewhat higher metabolic energy when they have other goals in mind. For example, a subject may aim to walk so as to minimize the effects of any number of other characteristics of a movement, such as: pain (the movement is carried out to keep pain low), risk of an adverse event such as falling (the movement is carried out to keep low the probability of falling), cognitive load (the subject is carrying out a cognitive task while performing the movement, such as walking while talking), etc. In addition, subjects have the freedom execute a movement with various speeds (e.g. faster or slower walking speed), or various trajectories (e.g. longer or shorter step lengths). We therefore require a model of human movements to be able to include further goals or principles in addition to energy minimization. We expect the effects of these other goals or principles on movement patterns to be interesting subjects of study in their own right, and of greater importance than energy minimization in the study movement in the contexts of aging, locomotor rehabilitation, movement disorders, etc.

In this project, we develop a computational formalism to describe the process by which a subject selects a particular movement trajectory to attain a specific goal. We expect the model to account for the observed human movement patterns as resulting from patterns in how movements are selected. The formalism that we develop is naturally open to the inclusion of further elements intended to account for the effects of other factors or goals on how movements are executed. We carry out this project out in three parts. In the first part (Sec. 2), we lay out the movement utility formalism within which we can develop models to study various kinds of movements together with a metabolic energy formalism for developing models to estimate the metabolic energies of particular movements. In the second part (Sec. 3), we develop a model of the metabolic energies of normal walking gaits using the metabolic energy formalism and plug this model into the movement utility formalism to construct a model to describe the selection of normal walking gaits. We validate this model in two stages, first we show that the metabolic energy model can be fit well to the available empirical data in Atzler & Herbst [12] and use this fit to provide empirical values to free parameters in the model, then we show that the movement utility model predicts selection of normal walking gaits that have the relationship between avg. walking speed and avg. step length observed by Grieve. [29] In the third part (Sec. 4), we propose a physical mechanism for how a subject might estimate the metabolic energy of a movement.

**2 Modeling Movement Selections**

We begin with the first part of the project in which we develop the movement utility and metabolic energy formalisms to describe movement selection. In the approach that we develop, we suppose that the subject selects the optimal movement out of a range of feasible movements. Mathematically, we express optimality by constructing a utility measure that can be calculated for all feasible movements in the range, and finding the movement of highest utility. As the utility includes a component of metabolic energy minimization, we also include a simple model of net metabolic energy that seems able to provide a reasonably good fit to empirical data (we use the term *metabolic energy* to refer to the metabolic energy above basal metabolic rate throughout).

*2.1 Segment Models*

We describe the pose of the body at time $t$ using a segment model consisting of a number of rigid segments with endpoints attached to each other. We define a *joint* to be points where segments attach to each other or endpoints of segments where those endpoints do not attach to another segment. As an example, the segment model given in [10] consists of 14 segments: (i) head, (ii) trunk, (iii) upper arms, (iv) lower arms, (v) hands, (vi) thighs, (vii) legs, and (viii) feet. We cause the segment model to execute a movement by specifying the joint-trajectories in space of the $n$ joints. We assume these joint-trajectories are constrained to produce physically sensible movements of the body.

*2.2 Movement Parameters*

As we aim to construct a movement utility model to describe movement selection, we would like a convenient way to search among all the feasible movements in the range. We do this by approximating the true joint-trajectories with joint-trajectories $\vec{x}_n(t)$ that are determined by a finite set of parameter values. The intent is that the range of feasible movements is replaced by a range of parameter values and that any particular set of parameter values determines the movement that is executed. Thus the movement utility problem becomes one of search the space of parameter values for those that maximize the utility of the movement. The parameter values are particular values of a set of *movement parameters* $\xi_1, \ldots, \xi_m$ defined on the range of parameter values. We can think of a range of movements as being described by a single, general set of trajectories $\vec{x}_n(t|\xi_1, \ldots, \xi_m)$ which are functions of the movement parameters so that the movement that is executed changes continuously as the movement parameters are changed continuously with each set of values of the movement parameters corresponding to particular set of parameter values.

To illustrate what we intend by the movement parameters, we can look at the case of walking gait. A simple model of walking, such as used in [29, 30], characterizes how the walk is carried out using the gait parameters of avg. walking speed and the avg. step length. More detailed models may be constructed by including other, measurable, aspects of walking such as: (i) avg. change in height of the torso over the course of a step, (ii) avg. height the swing foot is lifted from the ground during a step, or (iii) avg. fraction of a step spent in double support. Thus, depending on the detail in which walking is to be studied we can parameterize the range of possible walking gaits using the gait parameters: avg. walking speed, avg. step length, avg. change in height of the torso, etc. The intent is that the movement parameters suffice to parameterize all the approximate joint-trajectories $\vec{x}_n(t)$ in the movement, and that they can be measured by observation of the movement. We expect that, as more movement parameters are included in the model, the approximate joint-trajectories better approximate the true joint-trajectories.

*2.3 Utility and Goal Functions*

Once we have defined the movement parameters that determine the trajectories of the joints, the problem of selecting a movement becomes one of finding the values of the movement parameters that correspond to the maximum utility movement. The utility of a movement measures how well the movement satisfies a number of other factors or goals — these may include things like keeping pain low, keeping low the risk of an adverse event such as falling, compensating for cognitive load or dual-tasking, etc., as well as preferences in how to execute the movement. In the example of walking, goals may include: (i) walking with faster or slower avg. walking speed, or (ii) walking with longer or shorter avg. step length. We measure the utility of a movement using a *utility function* $J(\xi_1, \ldots, \xi_m)$. We construct the utility function as the weighted sum of a set of *goal functions* $G_i(\xi_1, \ldots, \xi_m)$:

$$J\left(\xi_1, \ldots, \xi_m\right) = \sum_{i=0}^{\mu} \lambda_i G_i\left(\xi_1, \ldots, \xi_m\right). \tag{1}$$

The goal function $G_i(\xi_1, \ldots, \xi_m)$ provides a measure of how well a movement meets the *i-th* goal. The *weights* $\lambda_i$ reflect the importance the subject has elected to place on meeting the associated goals. Goal functions are unique up to addition and/or multiplication by a constant. Addition by a constant simply introduces a constant term into the utility function which will have no effect on the maximization process that finds the movement of maximum utility, while multiplication by a constant can be compensated for by simply rescaling the associated weight.

*2.4 Metabolic Energy Model*

We single out for more detailed analysis the goal of minimizing the metabolic energy of the movement; we define this goal to be $G_0(\xi_1, \ldots, \xi_m)$. We denote the metabolic energy of the movement by $W(\xi_1, \ldots, \xi_m)$, and we express the goal of minimizing the metabolic energy using $G_0(\xi_1, \ldots, \xi_m) = W(\xi_1, \ldots, \xi_m)$; the utility $J(\xi_1, \ldots, \xi_m)$ becomes:

$$J\left(\xi_1, \ldots, \xi_m\right) = -\lambda_0 W\left(\xi_1, \ldots, \xi_m\right) + \sum_{i=1}^{\mu} \lambda_i G_i\left(\xi_1, \ldots, \xi_m\right). \tag{2}$$

Let us suppose the body is moving with joint-trajectories $\vec{x}_n(t|\xi_1, \ldots, \xi_m)$. We denote by $\vec{F}_n(t|\xi_1, \ldots, \xi_m)$ the net effect of the muscle forces on the joints and by $\vec{v}_n(t|\xi_1, \ldots, \xi_m) = \dot{\vec{x}}_n(t|\xi_1, \ldots, \xi_m)$ the velocities of the joints. We note that in the presence of external forces such as gravity, the forces $\vec{F}_n(t|\xi_1, \ldots, \xi_m)$ may not be the same as the net forces acting to move the body. We assume that we can approximate the metabolic rate $\dot{W}(t)$ for the movement as the sum of the metabolic rates $\dot{W}_n(t)$ calculated locally at each joint:

$$\dot{W}(t) \approx \sum_{n=1}^{N} \dot{W}_n(t). \tag{3}$$

The metabolic rate $\dot{W}_n(t)$ can be decomposed into a sum of two metabolic rates: the metabolic rate $\dot{W}_n^F(t)$ of generating the muscle forces, and (ii) the metabolic rate $\dot{W}_n^E(t)$ of adding /removing mechanical energy to/from the body over the course of the movement (i.e. the metabolic rate associated with generating a mechanical power). [10] We suppose that the metabolic rate $\dot{W}_n^F(t)$ is a function of the muscle forces alone, and the metabolic rate $\dot{W}_n^E(t)$ is a function the muscle forces and velocity of the joint (giving it a close relationship to the mechanical power). The metabolic rate associated with a joint is therefore given by a function:

$$\dot{W}_n(t) = \dot{W}_n^F\left(\vec{F}_n(t)\right) + \dot{W}_n^E\left(\vec{F}_n(t), \vec{v}_n(t)\right). \tag{4}$$

For notational convenience, we have suppressed the movement parameters. The metabolic rate of the movement is the sum of the metabolic rates of the joints.

In the construction of the model used in this paper we have supposed that we can model the metabolic rate $\dot{W}_n(t)$ using functions of the muscle forces $\vec{F}_n(t)$. We [31-33] and others [7, 34-36] have proposed that humans select the trajectory or orbit of a movement by finding the orbit which minimizes the jerks $\dddot{x}_n(t)$ of the movement. In the present construction, we avoid calculating the orbits by simply choosing simple orbits that approximate the physical movements the model is describing. However, we expect the jerks $\dddot{x}_n(t)$ to bear some relationship to the rates of change of the muscle forces $\dot{\vec{F}}_n(t)$, and it has been argued in [7] that $\dot{\vec{F}}_n(t)$ may contribute to the metabolic rate. There may therefore be some relationship between the utility model with metabolic energy minimization that we construct here, and a utility model with jerk minimization. In the present treatment we suppose that we can neglect any contribution of the $\dot{\vec{F}}_n(t)$.

We do not have an explicit physical model for $\dot{W}_n^F\left(\vec{F}_n(t)\right)$, so we assume that we can approximate the (unknown) physical model by using a Taylor series expansion. We impose the physical requirements that $\dot{W}_n^F > 0$ (muscle forces always have a positive metabolic rate), and $\dot{W}_n^F = 0$ when $\vec{F}_n = 0$ (the metabolic rate for muscle activation is zero when the muscle is inactive). We find that the series expansion has the form:

$$\dot{W}_n^F\left(\vec{F}_n(t)\right) = \vec{F}_n^{\mathrm{T}}(t) \mathrm{E}_n \vec{F}_n(t) + \left[higher\ order\ terms\right]. \tag{5}$$

The lowest order term in (5) provides part of the model used to describe movement trajectories in [31]. Similarly, we do not have an explicit physical model for $\dot{W}_n^E\left(\vec{F}_n(t), \vec{v}_n(t)\right)$, so we again assume that we can approximate the (unknown) physical model by using a Taylor series expansion. We impose the physical requirements that $\dot{W}_n^E = 0$ when either $\vec{F}_n = 0$ or $\vec{v}_n = 0$ (the metabolic rate for adding/removing mechanical energy is zero when the mechanical energy is not changing). The latter requirement means the expansion only contain terms containing both $\vec{F}_n(t)$ and $\vec{v}_n(t)$. Suppressing the movement parameters, the expansion has the form:

$$\dot{W}_n^E\left(\vec{F}_n(t), \vec{v}_n(t)\right) = \vec{F}_n^{\mathrm{T}}(t) \mathrm{H}_n \vec{v}_n(t) \\ + \left[higher\ order\ terms\right]. \tag{6}$$

The quantities $E_n$ ("epsilon sub n") and $\mathrm{H}_n$ ("eta sub n") are constant matrices such that, since the metabolic rates are expected to be positive, the products $\vec{F}_n^{\mathrm{T}}(t)E_n\vec{F}_n(t)$ and $\vec{F}_n^{\mathrm{T}}(t)\mathrm{H}_n\vec{v}_n(t)$ give non-negative scalar values. We allow that the matrices $\mathrm{H}_n$ may take on different values when mechanical energy is added to or removed from a segment, though we require they be constant in each case. As we might have expected, the lowest order term in the expansion (the first term on the right-hand side (RHS)) of (6) is bears a close resemblance to the mechanical power. We assume we can get a reasonable approximation the metabolic rate by keeping only the lowest order term on the RHS in each of (5) and (6).

*2.5 Discussion*

On the one hand, human subjects have complete freedom in how they execute a movement; on the other, they do appear to normally exhibit patterns in how they execute movements. The model of movement selection we have laid out is able to provide an account of the both the complete freedom subjects have in how they execute a movement and the patterns subjects typically exhibit when selecting movements. The model accounts for the observed movement patterns by allowing that there is a pattern in the other factors or goals that the movement is intended to meet and therefore a pattern in how movements are selected. As the goals for the movement reflect physical circumstances (both external and biomechanical) under which the movement is made, as well as subject preferences, movement patterns are a result of regularity of physical circumstances and personal preferences. The model can be easily made to accommodate goals of interest to the study and characterization of movements in relationship to aging, locomotor rehabilitation, movement disorders and so on; some examples of such goals are: keeping pain low, keeping low the risk of an adverse event such as falling, compensating for cognitive load or dual-tasking, etc. The presence of these goals will have the effect of producing a movement pattern, albeit one that differs from the movement pattern for subjects not having these goals.

The development of a utility model for more clinically interesting goals requires the researcher to provide a computational model for an associated goal function. We do not expect the formulation of these goal functions to be trivial nor have we provided any guidance on how to construct them, but we also do not expect goal functions to be formulated as ad hoc mathematical formulas aiming to provide a description of movement on a case-by-case basis. We expect that, were the approach that we have presented followed through as a research program, principles for the formulation of the goal functions would emerge and they exist within a clearly articulated body of theory providing insight into human motor behavior.

**3 Selection of Normal Walking Gaits**

We move on to the second part of the project in which we illustrate how the formalism developed in Sec. 2 can be applied in practice by using it to guide the construction a movement utility model of the selection of normal walking gaits (i.e. the middling speed walking gaits selected by healthy young adults and adults under typical conditions and not the very slow walking gaits exhibited by persons with mobility problems or very fast walking gaits with speeds in the region where running becomes preferable). We proceed broadly following an approach along the lines of [9, 30] — we first provide an account of the observed metabolic energy of walking in Atzler & Herbst [12] and then we use this to provide an account of the pattern of normal walking gaits observed by Grieve. [29] The construction of the movement utility model for selection of normal walking gaits develops along the following stages: (i) we give a kinematic model and identify the gait parameters (Sec. 3.2), (ii) we give the general form of the utility function (Secs. 3.2 and 3.3), (iii) we derive the model for the metabolic energy of one stop during walking (Sec. 3.4), (iv) we use available empirical data in Atzler & Herbst [12] to estimate values for the free parameters in the metabolic energy model (Sec. 3.5), (v) we give models for the other goals in the utility function (Sec. 3.6), (vi) we select the walking gait by maximizing the utility function and find it is a function of the weights of the goals (Sec. 3.7), and (vii) we provide an account of the pattern of normal walking gaits observed by Grieve [29] using regularities in how the weights of the goals affect the selected normal walking gait (Sec. 3.8). We find it contributes to the clarity of how the metabolic and mechanical energies relate to express metabolic energy in calories and mechanical energy in joules; these two measures of energy are related to each other by 1.0 cal = 4.2 J; we convert to common units to calculate dimensionless energy efficiencies.

*3.1 Some Anthropometric Values*

For convenience, we now give a number of anthropometric values relevant to the calculations in this section. A subject with mass $M$ and height $H$ has a mass in each leg (i.e. thigh, shank, and foot) of about $\mu = rM$, and the length of the leg of about $L = \rho H$ where $r = 0.16$ and $\rho = 0.53$. [10] The mass of the torso carried by the stance leg during a step is $m = (1 - 2r)M$. The avg. walking speeds and step lengths for adults aged 20-49 years are $v° \approx 1.3$ m·s$^{-1}$ and $s° \approx 0.61$ m. [37] The gross metabolic rate during standing is about 0.31 cal·kg$^{-1}$·s$^{-1}$, and during walking it becomes about 0.50 to 1.7 cal·kg$^{-1}$·s$^{-1}$ for avg. walking speeds of 0.2 m·s$^{-1}$ to 1.9 m·s$^{-1}$. [38]

*3.2 Kinematic Model*

We use a two segment model with one segment for each leg — the legs are straight and do not bend at the knee. The mass of the torso is located at the *torso* which is the point where the two segments meet; the mass of each leg is located in the feet at the far end of the leg segments from the torso. The model therefore has three joints: (i) torso, (ii) left foot, and (iii) right foot. We require that the torso maintain a constant height throughout the walk and maintain a constant speed along a straight line parallel to the ground. During one step, one leg is the *stance leg* which supports the torso as the torso moves over it, and the other is the *swing leg* which swings under the torso; the feet of the two legs are the *stance foot* and *swing foot*, respectively. The stance foot remains fixed on the ground while the swing foot glides a negligible distance above the ground; the legs lengthen or shorten as needed by the movement — we intend the amount of lengthening and shortening to be consistent with reasonable knee-bending during walking.

We only look at steady state walking gaits that are in progress and maintain constant values for the gait parameters; we do not consider the process of starting or stopping a walking gait. We describe walking gait using two gait parameters: (i) the avg. walking speed (the constant speed of the torso) $\xi_1 = v$, and (ii) the avg. step length (the distance between the feet when they are both on the ground) $\xi_2 = s$ (the avg. stride length, or the distance the foot of the swing leg moves during a step, is twice the avg. step length). We define the unit vector $\hat{v}$ to be the direction of motion of the torso. The gait parameter $v$ giving the avg. walking speed should not be confused with the velocities $\vec{v}_n$ appearing in the metabolic energy model or the unit vector $\hat{v}$; the diacritical mark or the absence of one suffices to distinguish them.

We require the torso to move horizontally with constant velocity. We use a simple model to describe the motion of the swing leg in which the swing leg moves symmetrically so that the swing foot glides horizontally, a negligible distance above the ground, with a constant acceleration during the first half of the step and a constant deceleration during the second half. Mathematically, the motions of the torso and swing foot are:

$$\begin{aligned}\dot{\vec{x}}_{torso}(t) &= v\hat{v}, \\ \ddot{\vec{x}}_{foot}(t) &\approx \begin{cases} (8v^2/s)\hat{v}, & 0 \leq t \leq s/2v, \\ -(8v^2/s)\hat{v}, & s/2v < t \leq s/v. \end{cases}\end{aligned} \quad (7)$$

The movement of the swing foot in (7) is arguably the simplest movement with non-zero horizontal force; we discuss this this choice of swing foot movement in Sec. 3.9.

*3.3 Utility Function*

We assume a subject selects the optimal walking gait by selecting the walking gait that maximizes the utility for each step rather than for a larger number of steps; we discuss this further in Sec. 3.9. We construct a model where walking gait is selected to best meet three goals: (i) minimizing the metabolic energy per step $W(v,s)$, (ii) preferring to have a faster or slower avg. walking speed as expressed by a goal function $G_1(v)$, and (iii) preferring to have a longer or shorter avg. step length as expressed by a goal function $G_2(s)$; the utility function has the form:

$$J(v,s) = -\lambda_0 W(v,s) + \lambda_1 G_1(v) + \lambda_2 G_2(s). \quad (8)$$

*3.4 Metabolic Energy Model*

We now work out the functional form of the metabolic energy per step $W$. We suppress explicit mention of the gait parameters in function arguments to keep the notation compact. The healthy human body has been shown to conserve mechanical energy during walking, [4, 39] so some of the mechanical energy in a step has been carried over from the previous step. We assume that the metabolic energy expended to restore the mechanical energy that has been lost from the previous step is approximately constant over the range of walks, so $\sum_{n=1}^{3} W_n^E \approx W_0$ (this value must be measured empirically). We assume that the metabolic energy associated with holding the body up against gravity is approximately constant over the range of walking gaits and is the same as that for standing so that we can treat it as part of the background metabolic rate that we ignore in the calculation of the metabolic energy of a step. All that remains is to calculate the metabolic energy associated with generating the muscle forces $\sum_{n=1}^{3} W_n^F$.

We denote the muscle force applied by the stance leg to the torso by $\vec{F}_{st}(t)$ and the force applied by the torso to the swing leg by $\vec{F}_{sw}(t)$; we assume the stance foot is fixed on the ground during the step. The matrices describing the metabolic rate of generating muscle force are denoted by $E_{st}$ and $E_{sw}$, respectively. The time required to execute a step is $T = s/v$; the metabolic energy per step takes the form:

$$\begin{aligned}W &\approx W_0 + \int_0^{s/v} \vec{F}_{st}^{\mathrm{T}}(t) \mathrm{E}_{st} \vec{F}_{st}(t) dt \\ &+ \int_0^{s/v} \vec{F}_{sw}^{\mathrm{T}}(t) \mathrm{E}_{sw} \vec{F}_{sw}(t) dt.\end{aligned} \quad (9)$$

We look first at the muscle forces $\vec{F}_{st}(t)$ applied by the stance leg to the torso as illustrated in Fig 1. We require the torso to move horizontally with constant velocity $\dot{\vec{x}}_{torso}(t)$ in (7). During the first half of the step, gravity pulls the torso in the direction $-\hat{v}$, while, during the second half, gravity pulls the torso in the direction $\hat{v}$. As we have assumed that $W_0$ accounts for any loss of mechanical energy, the torso already has the mechanical energy needed to carry it forward with the required speed, the muscle forces counteracts the force of gravity.

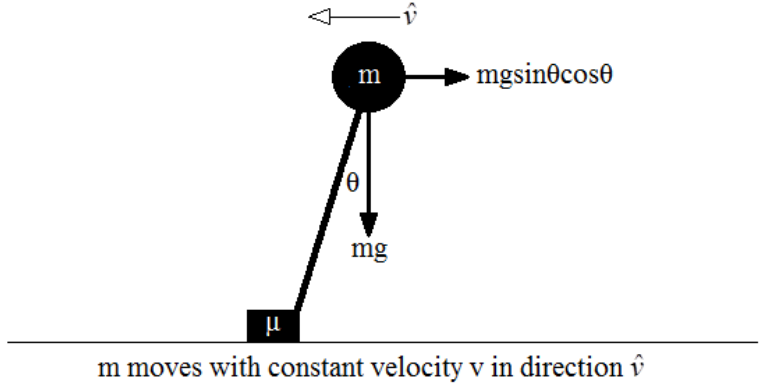

**Figure 1.** The motion of the torso over the stance leg during one step of normal walking gait. The body moves over the stance leg at a constant speed $v$. To maintain the constant speed, the muscles must provide force to compensate for the effect of gravity on the body's speed.

The effect of gravity on the torso is determined by the angle $\theta(t)$ of the stance leg with respect to the vertical. We define $\theta(t)$ so that it is negative during the first half of the step, and positive during the second half. The horizontal force related to the torque on the inverted pendulum by gravity is $mg\sin\theta\cos\theta = (1/2)mg\sin 2\theta$ where $g$ denotes the acceleration of gravity; we find $\theta(t)$ and $\vec{F}_{st}(t)$ satisfy:

$$L \sin\theta(t) = vt - s/2,$$
$$\vec{F}_{st}(t) = (1/2) mg \sin 2\theta(t) \hat{v}. \tag{10}$$

We note that, when observed over several steps, the velocity of the torso is constant, while the force $\vec{F}_{st}(t)$ acting on the torso is discontinuous at toe-off/heel strike.

We look next at the muscle forces $\vec{F}_{sw}(t)$ applied by the torso to the swing leg as illustrated in Fig. 2. We require the swing foot to move horizontally with acceleration $\ddot{x}_{foot}(t)$ in (7). During one step, the body moves a distance of one avg. step length, and the swing foot travels a distance of one stride length horizontally in direction $\hat{v}$. The swing leg begins and ends the swing at rest, and so must accelerate and decelerate as required over the course of the swing. During the first half of the step, gravity pulls the swing leg in the direction $\hat{v}$, while, during the second half, gravity pulls the swing leg in the direction $-\hat{v}$. We have noted that the body acts to conserve mechanical energy, so much of the mechanical energy in the swing leg has been taken over from the previous swing of the other leg, and $W_0$ accounts for any further mechanical energy that must be added or removed, thus, the muscle forces must simply generate the required acceleration and deceleration while counteracting the force of gravity.

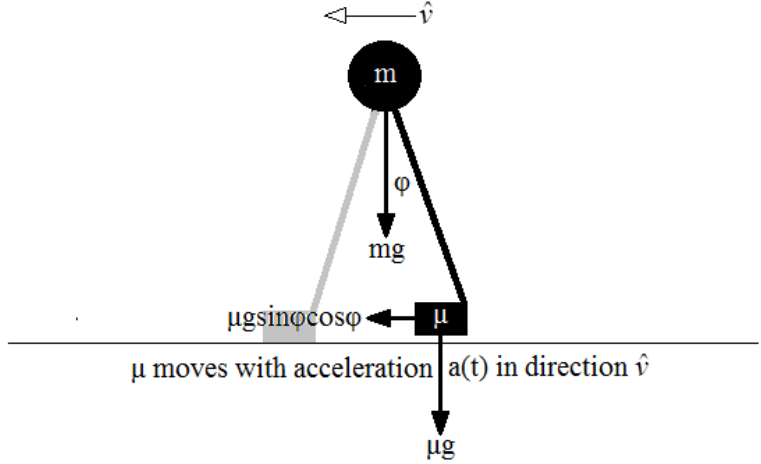

**Figure 2.** The motion of the swing leg under the torso during one step of walking gait. The swing leg moves symmetrically under the body, accelerating with an acceleration $\ddot{x}(t)$ during the first half of the step, and decelerating with an acceleration $-\ddot{x}(t)$ during the second half. The muscles must provide the force that generates the acceleration and deceleration and compensate for the effect of gravity on the swing leg.

The effect of gravity on the swing leg is determined by the angle $\varphi(t)$ of the swing leg with respect to the vertical. We define $\varphi(t)$ so that it is negative during the first half of the step, and positive during the second half. As the acceleration of the swing leg is symmetric through the step, we can calculate the metabolic energy of the swing leg by doubling the metabolic energy for the first half of the step. The horizontal muscle forces that must be applied to the leg to generate the constant acceleration in (7) during the first half of the step is $\mu a + \mu g \sin\varphi\cos\varphi = \mu a + (1/2)\mu g \sin 2\varphi$; we find $\varphi(t)$ and $\vec{F}_{sw}(t)$ for the first half of the step satisfy:

$$L\sin\varphi(t) = \left(4v^2/s\right)t^2 - vt - s/2, \quad 0 \leq t \leq s/2v,$$
$$\vec{F}_{sw}(t) = \left(\left(4\mu v^2/s\right) + \left(\mu g/2\right)\sin 2\varphi(t)\right)\hat{v}. \tag{11}$$

We note that, when observed over several steps, the speed of a given foot is continuous, while the force $\vec{F}_{sw}(t)$ acting on the foot is discontinuous at toe-off and heel strike, and the acceleration $\ddot{\vec{x}}_{foot}(t)$ is discontinuous at toe-off, heel strike, and midswing.

We define the constant parameters $\varepsilon_{st} = \hat{v}^T E_{st}\hat{v}$ and $\varepsilon_{sw} = \hat{v}^T E_{sw}\hat{v}$ (these values must be measured empirically). For small avg. step lengths $s$, we may take $\sin 2\theta \approx 2\sin\theta$ and $\sin 2\varphi \approx 2\sin\varphi$. The metabolic energy of the swing leg acting on the torso for a single step is:

$$\int_0^{s/v} \vec{F}_{st}^T(t) E_{st} \vec{F}_{st}(t)\,dt \approx \left(\varepsilon_{st} m^2 g^2 / 12L^2\right)s^3/v. \tag{12}$$

The metabolic energy of moving the swing leg for a single step is:

$$\int_0^{s/v} \vec{F}_{sw}^{\mathrm{T}}(t) \mathrm{E}_{sw} \vec{F}_{sw}(t) dt = 2\int_0^{s/2v} \vec{F}_{sw}^{\mathrm{T}}(t) \mathrm{E}_{sw} \vec{F}_{sw}(t) dt$$
$$\approx \left(16\varepsilon_{sw}\mu^2\right)v^3 / s - \left(10\varepsilon_{sw}\mu^2 g / 3L\right)vs + \left(\varepsilon_{sw}\mu^2 g^2 / 5L^2\right)s^3 / v. \tag{13}$$

*3.5 Empirical Study I (Atzler & Herbst, 1927)*

We now make the metabolic energy model developed in (9), (12), and (13) into an estimator of the metabolic energy during walking over the range of allowed walking gaits by using empirical data to produce estimates for the parameter values $W_0$, $\varepsilon_{st}$, and $\varepsilon_{sw}$.

Atzler & Herbst [12, 40] observed one subject (male, $M = 68$ kg, $H = 1.7$ m, aged 39 years, mass-normalized resting metabolic rate $\dot{W}_{rest}/M = 0.30$ cal·kg$^{-1}$·s$^{-1}$) perform a variety of walking gaits, and measured the metabolic energy for each walking gait using a Zuntz-Geppert respiratory apparatus and the Douglas Bag Technique. The subject was trained to walk on a horizontal treadmill using all 20 combinations of 4 avg. step lengths $s$ (0.46 m, 0.60 m, 0.76 m, and 0.90 m) and 5 avg. cadences $v/s$ (0.83 step·s$^{-1}$, 1.25 step·s$^{-1}$, 1.7 step·s$^{-1}$, 2.2 step·s$^{-1}$, and 2.5 step·s$^{-1}$); four trials were conducted for each distinct walking gait.

Combining (9), (12), and (13), the metabolic energy per step can be written as:

$$W \approx W_0 + \left(m^2 g^2 s^3 / 12 L^2 v\right)\varepsilon_{st}$$
$$+ \left(16\mu^2 v^3 / s - 10\mu^2 gvs / 3L + \mu^2 g^2 s^3 / 5L^2 v\right)\varepsilon_{sw}. \tag{14}$$

Using ordinary least-squares regression, this model fit the data for Atzler & Herbst's subject with $R^2 = 0.99$ and $p < 0.0001$, using estimated parameter values of:

$$\begin{aligned}
W_0 &\approx 9.0\, cal, \\
\varepsilon_{st} &\approx 2.5 \times 10^{-3}\, cal \cdot N^{-2} \cdot s^{-1}, \\
\varepsilon_{sw} &\approx 1.7 \times 10^{-3}\, cal \cdot N^{-2} \cdot s^{-1}.
\end{aligned} \tag{15}$$

Inspection of the 95% confidence intervals showed that all 3 parameters were statistically significant; the fit is shown in Fig. 3.

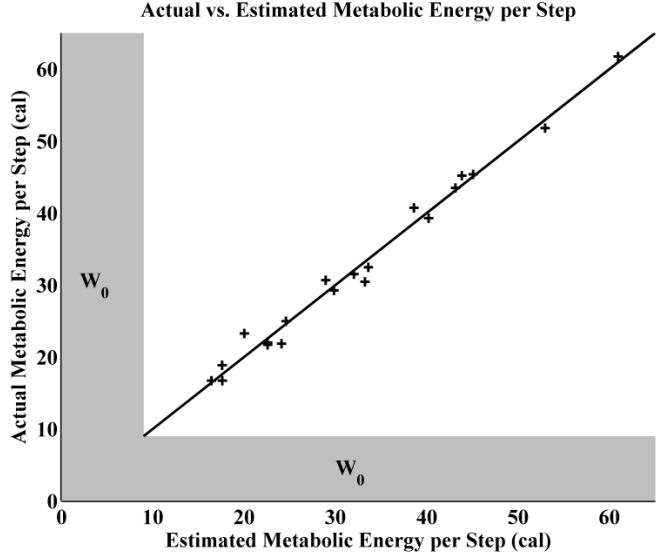

**Figure 3.** Actual vs. estimated metabolic energy per step. We use the model in (14) to fit the observed metabolic energies for the 20 walking gaits in Atzler & Herbst; the fit for the model has $R^2 = 0.99$ and $p < 0.0001$. The value for the constant parameter $W_0$ is indicated. For reference, we show a segment of the line with slope one passing through the origin.

*3.6 Other Goals*

The utility function in (8) includes two other goals: (i) the goal $G_1(v)$ of preferring to have a faster or slower avg. walking speed, and (ii) the goal $G_2(s)$ of preferring to have a longer or shorter avg. step length. Human walking gait appears to be described by Weber's law, [41] suggesting that the goal functions should be related to the perceived stimuli $\psi_v$ and $\psi_s$ associated with the avg. walking speed and avg. step length, respectively, where these stimuli follow Weber's law (i.e. $\psi_v = log(v/v_0)$ and $\psi_s = log(s/s_0)$ where $v_0$ and $s_0$ are constants with units of walking speed and step length, respectively). We assume the subject is aware of the stimuli associated with the two gait parameters rather than the gait parameters themselves and use goal functions that are equal to the stimuli: $G_1(v) = \psi_v$ and $G_2(s) = \psi_s$.

*3.7 Selecting a Normal Walking Gait*

The subject selects the normal walking gait that maximizes the utility function. Thus, the selected normal walking gait is the solution to the system of equations:

$$\begin{aligned} v\frac{\partial W}{\partial v}(v,s) &= \lambda_1 v_0 / \lambda_0, \quad (A) \\ s\frac{\partial W}{\partial s}(v,s) &= \lambda_2 s_0 / \lambda_0. \quad (B) \end{aligned} \quad (16)$$

We can write the metabolic energy model developed in (9), (12), and (13) compactly in the form $W(v,s) \approx W_0 + \alpha s^3/v - \beta vs + \gamma v^3/s$ where $\alpha$, $\beta$, and $\gamma$ are constants for the subject. We can now rewrite (16) as:

$$\begin{aligned}\alpha s^4 + \beta v^2 s^2 - 3\gamma v^4 &= -\left(\lambda_1 v_0 / \lambda_0\right) vs, \quad (A) \\ 3\alpha s^4 - \beta v^2 s^2 - \gamma v^4 &= \left(\lambda_2 s_0 / \lambda_0\right) vs. \quad (B)\end{aligned} \qquad (17)$$

The selected normal walking gait $v^* = v(\lambda_0, \lambda_1, \lambda_2)$ and $s^* = s(\lambda_0, \lambda_1, \lambda_2)$ lies at the intersection of the curves given by the equations in (17) and solves both equations simultaneously. We find that the values of the weights $\lambda_0$, $\lambda_1$, and $\lambda_2$ determine the normal walking gait.

We illustrate the curves A and B corresponding to the two equations in (17) in Fig. 4. We have used the height-normalized avg. stride length rather than the avg. step length in Fig. 4 to facilitate comparison to the corresponding figure in Grieve. [29] The subject selects the walking gait with avg. walking speed $v^*$ and avg. step length $s^*$. The values for the parameters $\alpha$, $\beta$, and $\gamma$ are those for Atzler & Herbst's subject; the values for the parameters $\lambda_1 v_0/\lambda_0$ and $\lambda_2 s_0/\lambda_0$ have been chosen so that the curves would intersect at a walk with height-normalized avg. walking speed and stride length $v^*/H \approx 1.0$ s$^{-1}$ and $2s^*/H \approx 1.0$ so that the selected walking gait would lie in the envelope of the walking gaits observed in Grieve. We have used contour lines to indicate constant metabolic energies per step in the metabolic energy landscape given by the model developed in (9), (12), and (13). The typical adult normal walking gait is indicated with height-normalized avg. walking speed v°/$H$ and avg. stride length 2s°/$H$.

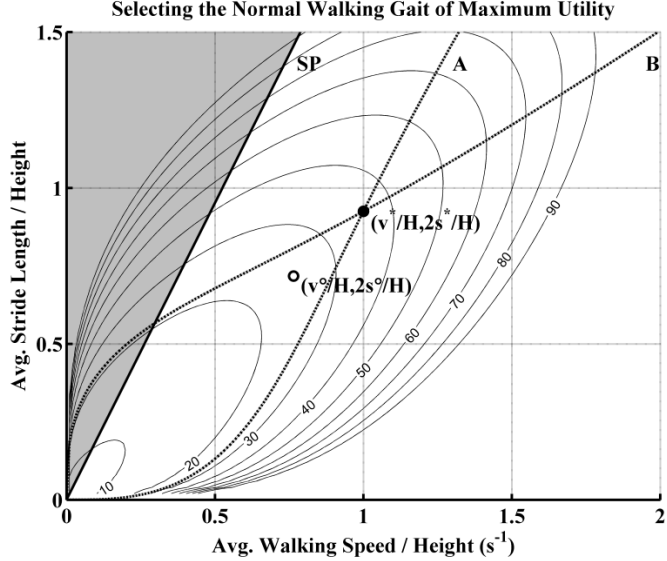

**Figure 4.** Selecting the normal walking gait of maximum utility. We show examples of the curves A and B in (17). The intersection of the two curves correspond to the walking gait of maximum utility; this walking gait has height-normalized avg. walking speed $v^*/H$ and avg. stride length $2s^*/H$. The white region indicates the region of walking gaits. The contour lines correspond indicate constant metabolic energies per step (in calories) in the metabolic energy landscape. The curve SP indicates the model for a walk where the swing leg moves as a simple pendulum. The typical adult walking gait is indicated with height-normalized avg. walking speed $v°/H$ and avg. stride length $2s°/H$.

In Fig. 4, we also give the curve SP that indicates the relationship between the avg. stride length and the avg. walking speed in a model where the swing leg moves as a simple pendulum (i.e. the avg. walking speed and avg. stride length are selected so that the motion of the swing leg is allowed to be driven entirely by the force of gravity). In the white region to the right of the curve SP, the time taken to swing the swing leg is shorter than one-half of the period of the simple pendulum, while in the grey region to the left of the curve SP, the time taken to swing the swing leg is longer than one-half of the period of the simple pendulum. The muscles act oppositely on the swing leg in the white and grey regions so that, during the initial part of swing, in the white region the muscle forces act to make the leg swing faster than it would by gravity along and in the grey region they act to make it swing slower than it would by gravity alone. The walking gaits observed in Grieve appear to be consistent with the postulate that subjects do not normally select walking gaits in the grey region in Fig. 4, and thus with the postulate that the curve SP provides a boundary on the region of normal walking gaits.

*3.8 Empirical Study II (Grieve, 1968)*

The selected normal walking gait is given by $v^* = v(\lambda_0, \lambda_1, \lambda_2)$ and $s^* = s(\lambda_0, \lambda_1, \lambda_2)$, so the weights $\lambda_0$, $\lambda_1$, and $\lambda_2$ determine the normal walking gait. We can now look at how patterns in the weighting of the goals account for observed walking gait patterns.

Grieve [29] observed 13 subjects (6 male and 7 female, aged 25-31 years) perform a variety walking gaits, and measured gait parameter values for the walks from film, using an analyzing projector. Subjects walked barefoot with minimal clothing in front of a camera filming at 32 fps. They were prompted to walk at a variety of avg. walking speeds using instructions like "very slowly and relaxed," "a little faster" and so on until they were walking as fast as possible after about ten transits. Hip, knee, and ankle markers permitted the angulation of the lower limbs to be measured in the sagittal plane, as well as the avg. walking speed and avg. stride length.

Grieve found that the self-selected avg. walking speeds and avg. step lengths for his subjects indicated a walking gait pattern of $s \propto v^{\kappa}$ where $\kappa \approx 0.42$. The normal walking gaits observed by Grieve appear to lie in an envelope that approximately includes an avg. walking speed and stride length $v^*/H \approx 1.0$ s$^{-1}$ and $2s^*/H \approx 1.0$ at the extreme of longer stride length and $v^*/H \approx 1.0$ s$^{-1}$ and $2s^*/H \approx 0.85$ at the extreme of shorter stride lengths.

The model generates a walking gait pattern that agrees qualitatively with the observations in Grieve when we require that subjects select walking gaits approximately on the curve obtained in Fig. 4 by using the curve SP for avg. walking speeds below $v^*$, and the curve B for avg. walking speeds above $v^*$. The subjects select walking gaits by holding the weights $\lambda_0$ and $\lambda_2$ constant while varying the weight $\lambda_1$. Geometrically, varying $\lambda_1$ has the effect of sweeping curve A across the curve obtained by combining curves SP and B, changing the position of the point $(v^*, s^*)$. Recalling that, in Grieve's study, different walking gaits were elicited by giving the subjects instructions as to how fast two walk, we find that subjects approximately fixed $\lambda_0$ and $\lambda_2$ based on the conditions of the setup of the study, and translated the instructions into values of the weight $\lambda_1$ that is associated with the goal $G_1(v)$ or the preference of having faster or slower avg. walking speeds.

Finally, we looked at how well the curve for B in (17) accounts for normal walking gaits performed under Grieve's controlled laboratory conditions. Grieve observed walking gaits approximately in the envelope where: (i) avg. walking speeds were in the range 0.25 s$^{-1} \leq v/H \leq 1.5$ s$^{-1}$, and (ii) avg. stride lengths were in the range $0.85 \leq 2s^*/H \leq 1.0$ at the avg. walking speed $v^*/H \approx 1.0$ s$^{-1}$. We used the values for $\alpha$, $\beta$, and $\gamma$ estimated for Atzler & Herbst's subject and found the values for $\lambda_2 s_0/\lambda_0$ that yielded walking gaits with stride lengths $2s^*/H = 0.85$ and $2s^*/H = 1.0$ at the avg. walking speed $v^*/H \approx 1.0$ s$^{-1}$. We found that the curve for B in (17) yields an approximate relationship of $s \propto v^{\kappa}$ where $\kappa$ is approximately 0.77 (for $2s^*/H = 0.85$) and 0.55 (for $2s^*/H = 1.0$), close to the value of 0.42 reported by Grieve.

*3.9 Discussion*

We have developed a relatively simple model of the forces that drive the body during normal walking gait. A more sophisticated approach would include more details of how humans walk such as how a subject pushes off with the swing foot at the beginning of the swing (toe-off) or places the swing foot on the ground at the end of the swing (heel strike) as is done in [30, 42]. Effectively, the approach we take is to assume that, so long as the torso moves horizontally with a constant speed, the body must use the muscles to generate the forces needed to counteract the effect of gravity over the course of a step so as to allow the torso to maintain the constant speed. Thus, the model approximately indicates the net forces that must be generated over the course of a step while glossing over the details of how the body physically produces the actual forces. In addition, the model glosses over the actual mechanisms by which mechanical energy is conserved and lost from step to step. It therefore provides no information about how processes that bring about this energy conservation may become less efficient and thus require the subject to use more metabolic energy to account for the needed mechanical energy. We expect this to be a factor in understanding very slow walking gaits as well as walking gait in the context of various movement disorders.

We have given an equation of motion for the swing foot in (7) that is the simplest — having symmetric, constant accelerations through the motion. However, other motions of the swing foot could be chosen for the model. We chose the model in (7) in order to keep the mathematics of the model simple. We expect alternate models that provide more physically realistic motions by having more complicated time-courses of acceleration for the swing foot to result in calculations for the metabolic energy of generating muscle forces that do not differ too significantly from (13).

We have expressed, in Sec. 3.6, the observation in [41] that normal walking gaits appear to be described by Weber's law in the model by using goal functions in which both the avg. walking speed and avg. step length are described using Weber's law. However, for the model to provide an account of the observations in Grieve, [29] it is only necessary that the avg. step length be described by Weber's law. It is thus possible that the goal function for the avg. walking speed may take another form. If it is the case that we have made the wrong choice for this goal function, some observations that may point to a more correct form for this goal function can be found in [43].

In the utility function model that we developed to describe the selection of normal walking gaits, we assumed that the subject maximizes the utility on a step-by-step basis. We opted for the step-by-step approach as it allows the subject to optimize the normal walking gait while it is proceeding, rather than finding the optimal normal walking gait prior to commencing to walk. Thus the subject does not need to store information about various walks in any way or otherwise perform elaborate calculations, but simply needs to figure the optimal normal walking gait in the course of executing the walk. When optimizing on a step-by-step basis the subject can find the optimal normal walking gait by estimating the utility for each step and making appropriate changes in the normal walking gait of the next step with the aim of increasing the utility. In this way, we can look at normal walking gait as a sequence of iterations in an on-going optimization process.

**4 Physical Estimation of Metabolic Energy**

We conclude with the third part of the project in which we put forward a simple mechanism for how a subject estimates the metabolic energy of a movement physically. The movement utility formalism that we laid out in Sec. 2 allows subject to assign higher utilities to movements with lower metabolic energies. This presupposes that subjects have some physical means of estimating the metabolic energy of movements. We observe that by minimizing the muscle forces the subject can minimize the metabolic energy of generating muscle forces.

*4.1 Metabolic Energy and Muscle Forces*

We may write the metabolic energy $W^F$ of generating the muscle forces approximately as (cf. Sec. 2.4):

$$W^F \approx \varepsilon \int_0^T \vec{F}(t)^2 \, dt = \left(\varepsilon F_0^2\right)\left[\left(1/T\right)\int_0^T \left(\left|\vec{F}(t)\right|/F_0\right)^2 dt\right]T. \tag{18}$$

Here $F_0$ is a constant with units of force. We observe that the metabolic energy of the movement is proportional to the time-avg. of the square of the magnitude of the net applied force $\left|\vec{F}\right|$ multiplied by the movement time $T$. We therefore find that we can minimize the metabolic energy by minimizing this quantity.

In the static case where the body applied a force but the limb does not move, the perceived magnitude of muscle forces $\psi_F$ is related to the actual magnitude of the muscle forces $|\vec{F}|$ according to Stevens' power law and has the form $\psi_F \approx \left(|\vec{F}|/F_0\right)^{1.7}$. [44, 45] We can write this equivalently as the time-avg. of the constant applied force measured over a time $T$; this gives;

$$\psi_F \approx \left(1/T\right)\int_0^T \left(\left|\vec{F}(t)\right|/F_0\right)^{1.7} dt \qquad (19)$$

We assume that (19) holds in the dynamic case of movement where the magnitude of the muscle forces $|\vec{F}|$ is changing and over times typical movement times $T$. We note that (18) involves the time-avg. of $|\vec{F}|^2$ while (19) involves the time-avg. of $|\vec{F}|^{1.7}$. As the two time-avgs. are approximately equal in value, we replace the time-avg. in (18) with that in (19), and thereby express metabolic energy $W^F$ of generating the muscle forces in terms of the perceived magnitude of the muscle forces $\psi_F$:

$$W^F \approx \left(\varepsilon F_0^2\right)\psi_F T. \qquad (20)$$

Thus, the subject can minimize the metabolic energy $W^F$ of generating the muscle forces by minimizing the value $\psi_F T$ of the movement.

*4.2 Discussion*

The mechanism that we have proposed for minimizing the metabolic energy of a movement can be used to create utility functions in which higher utility movements are those that minimize the metabolic energy $W^F$ associated with generating the muscle forces rather than minimizing the metabolic energy $W = W^F + W^E$ (cf. Sec. 2.4). Thus the mechanism suffices to account for minimization of metabolic energy the model for the selection of walking gait that we have developed in Sec. 3 but fails to account for all the metabolic energy when external work is done (e.g. walking while pulling a cart). While, in general, minimization of $W^F$ may not result in movements with minimum $W$, we expect that minimum $W^F$ movements should result in movements with lower mechanical energies and, as a result, relatively low values $W$. We expect, therefore, that minimization of $W^F$ might serve as a heuristic to generate movements with relatively low metabolic energies.

**Acknowledgments**

This work was supported by supported by the National Science Foundation under Grants 1407928 and 1111965 and by the National Institute on Aging under Grants NIA P30AG024978 and 5RC1AG36121-2. We thank our colleagues from Northeastern University, Holly Jimison and Misha Pavel, who provided insight and expertise that greatly assisted the research, although they may not agree with all of the interpretations/conclusions of this paper.